\documentclass{article}

\usepackage[backend=biber,style=nature]{biblatex} 
\title{Why Has Advanced Commercial HVAC Control Not Yet Achieved Its Promise?}

\author{}
\author{Gregor P. Henze, University of Colorado Boulder, gregor.henze@colorado.edu \\
Kevin J. Kircher, Purdue University, kkirche@purdue.edu \\
James E. Braun, Purdue University, jbraun@purdue.edu}

\date{\today}

\begin{document}

\maketitle


\begin{abstract}
    Over the last two decades, research and development efforts have shown that advanced control of heating, ventilation, and air conditioning (HVAC) equipment in commercial buildings can improve energy efficiency, reduce emissions, and turn buildings into active participants in the power grid. Despite these efforts, advanced commercial HVAC control has not yet seen widespread adoption. In this paper, we argue that the research community can help companies deploy advanced HVAC control at speed and scale by reorienting research efforts toward clearly demonstrating the business case for adoption. To support this argument, we draw on findings from the 2023 Intelligent Building Operations Workshop, which brought together researchers, entrepreneurs, and representatives from industry and government to discuss current business offerings, state-of-the-art field demonstrations, barriers to adoption, and future directions.
    
\end{abstract}

\section{Introduction}

Advanced control of commercial heating, ventilation and air conditioning (HVAC) equipment has great potential to improve energy efficiency, reduce air pollution and greenhouse gas emissions, and turn buildings into active participants in power grid operations. The authors' recent review of field demonstrations of model predictive control (MPC) and reinforcement learning control (RLC) -- two popular algorithms for advanced commercial HVAC control -- studied 56 field tests reported in 36 peer-reviewed publications between 2005 and 2023 \cite{khabbazi}. The review found that both MPC and RLC can be expected to reduce costs and emissions by 15--25\% relative to conventional rule-based control. Scaled to the entire commercial building fleet, these savings could substantially reduce energy costs and greenhouse gas emissions. For example, in the United States, where HVAC systems in commercial buildings cause about 7\% of all greenhouse gas emissions \cite{epa}, 15--25\% savings could reduce the country's greenhouse gas emissions by 1--2\%, a comparable amount to all emissions from domestic aviation \cite{faa}, and save building operators \$13--22 billion per year \cite{cbecs}.

Despite its potential, advanced commercial HVAC control has not achieved widespread industry adoption. While major building automation companies have explored MPC, these efforts have largely ended. For example, to the best of our knowledge, Siemens discontinued its commercial HVAC MPC activities around 2016 after its OptiControl I and II projects \cite{sturzenegger2015model}, and Johnson Controls dissolved its commercial HVAC MPC team \cite{madison2015johnson, rawlings2018economic} around 2019. A number of start-up companies, several of which are discussed in this paper, offer MPC or RLC products, or related sensing or data integration technologies, but so far have not achieved significant market share.

We believe that few companies have adopted advanced commercial HVAC control because its business case has not been clearly demonstrated, is not an attractive offering today, or both. To assess project finance metrics (such as payback period, net present value, or internal rate of return), business decision-makers need to understand costs as well as benefits. Today, the research community understands the real-world benefits of advanced commercial HVAC control fairly well. However, the benefits may not have been convincingly conveyed to decision-makers in business, and more importantly, deployment costs are hardly understood at all. Very few deployment cost estimates can be found in past field studies, and the few available estimates are not encouraging. For example, Sturzenegger {\it et al.}  found after deploying MPC in five floors of an office building that ``the required initial investment is likely too high to justify the deployment in everyday building projects on the basis of operating cost savings alone'' \cite{sturzenegger2015model}. Similarly, Blum {\it et al.} reported that implementing MPC in two floors of an office building required 10 person-months of engineering labor \cite{blum2022field}.

The central argument of this paper is that the research community should reorient its efforts to emphasize characterization and improvement of the business case for advanced commercial HVAC control. Assuming finite resources for research and development, this will entail de-emphasizing the algorithmic refinements that currently dominate the research literature. Instead, this paper argues for a) synthesizing the lessons learned from past field studies; b) conducting more, and more focused, field studies; c) focusing as much on deployment costs as past studies have focused on economic and environmental benefits; and d)  studying how the costs and benefits of deploying advanced HVAC control depend on the building characteristics, climate, utility rate structure, and other boundary conditions. We believe that these efforts, if successful, will convince decision-makers in industry to invest in creating advanced HVAC control products and, ultimately, get the technology onto a learning curve similar to those that have radically reduced the costs of wind turbines, solar photovoltaics, and batteries over the last decade \cite{lazard}.

This viewpoint paper represents the distilled views of the participants in the 2023 Intelligent Building Operations IBO Workshop, one decade after a similar viewpoint paper in this journal by one of the authors asked whether MPC represented a quantum leap for building controls \cite{henze2013model}. This workshop was the sixth in a series, held alternately at the University of Colorado Boulder and Purdue University, that began in 2011. The 2023 workshop, titled ``Bridging the Abyss from Algorithms to Applications,'' focused on how researchers, engineers, and entrepreneurs can help move advanced HVAC control technology out of the laboratory and into real buildings at speed and scale. The participants hailed from start-up companies, large incumbent building technology companies, the United States Department of Energy, and from universities and national laboratories in the United States and Europe. The organizers of the workshop -- also the authors of this paper -- collectively have 90 years of research and development experience in the field of advanced HVAC control. The workshop and this paper grew in part out of our disappointment that the technology that we have worked on for so long still has not seen widespread adoption. 

This paper is organized as follows. To familiarize the reader with the current state of the advanced commercial HVAC control field, Sections \ref{businessOfferings} and \ref{fieldDemonstrations} provide brief samplings of current business offerings and recent state-of-the-art field demonstrations, respectively. Informed by the background information from Sections \ref{businessOfferings} and \ref{fieldDemonstrations}, Section \ref{futureDirections} then discusses challenges and opportunities related to business considerations (Section \ref{businessConsderations}), data access and system integration (Section \ref{dataAccess}), mandates for new business thinking (Section \ref{businessThinking}), and truth in advertising (Section \ref{truthInAds}). Section \ref{conclusion} concludes by reflecting critically on the state of the field and recommending new research directions. Detailed summaries of IBO workshop presentations can be found in the appendices.

\section{Examples of Today's Business Offerings}
\label{businessOfferings}

A handful of established companies dominate the market for automation and control systems in large commercial buildings. While these companies have participated in past IBO workshops, they were not well-represented in 2023. To the authors' knowledge, these companies are not currently devoting significant effort to integrating advanced commercial HVAC control algorithms such as MPC and RLC into their product offerings. For example, Siemens phased out their OptiControl project \cite{sturzenegger2015model} that was active around 2012--2016, and Johnson Controls disbanded their MPC team \cite{madison2015johnson, rawlings2018economic} that was active around 2014--2019. However, some smaller companies and start-ups now offer advanced HVAC control products. As a sample of current business activity in the field, the 2023 IBO workshop featured presentations from BuilTwins, Belimo Automation, Passive Logic, and QCoefficient.

BuilTwins \cite{builtwins}, a start-up that recently spun out of KU Leuven in Belgium, uses detailed physics-based modeling to develop high-fidelity representations of thermal and mechanical systems \cite{jorissen2021strengths}. BuilTwins envisions applying these digital twins to provide services such as design optimization, commissioning, performance monitoring, fault detection and diagnostics, and control \cite{jorissen2022white}. BuilTwins' HVAC control platform uses nonlinear MPC enabled by custom solvers and detailed Modelica models of all relevant building components \cite{wetter2014modelica}. Filip Jorissen, co-founder and chief technical officer, stated that BuilTwins has demonstrated aspects of their technology in several office buildings, care homes, and university buildings.

Belimo Automation's signature product, the Energy Valve, is an intelligent flow control device for HVAC systems with water-based thermal distribution \cite{belimo}. The Energy Valve consists of a controllable valve, a flow meter, and supply and return water temperature sensors, along with Internet-connected on-board processing. In addition to reducing energy use by avoiding valve operation with low temperature differences and high flow (and therefore high pumping energy), the Energy Valve provides real-time temperature and flow measurements. These measurements can facilitate building modeling and state estimation, key steps in system-wide supervisory control via MPC or RLC, by providing granular estimates of thermal demand. They can also facilitate fault detection and diagnostics by pinpointing any anomalous flow or temperature conditions that arise. The Energy Valve also facilitates building modeling and state estimation, key steps in system-wide supervisory control via MPC or RLC.

Passive Logic, a start-up that has procured about \$85 million in funding to date, has developed a completely new Building Automation System (BAS) that employs both novel hardware and software \cite{passivelogic}. Passive Logic uses automated data acquisition processes to develop digital twins of building systems using a graph-based representation of networked HVAC components \cite{harvey2022autonomous}. Passive Logic's technology can fit model parameters and optimize control actions by differentiating through the graph, similar to back-propagation in neural network training. Like BuilTwins, Passive Logic envisions using their digital twins and supporting software to provide a range of services, including control system design, deployment, commissioning, and monitoring.

QCoefficient \cite{qcoefficient} provides grid-interactive HVAC control for large commercial buildings, including several skyscrapers in Chicago and New York City. QCoefficient typically optimizes zone air temperature setpoints daily over a 24-hour prediction horizon using EnergyPlus models of buildings and HVAC equipment. QCoefficient can optimize for energy efficiency, energy costs, peak demand, annual capacity charges, greenhouse gas emissions, and/or indoor air quality. Vince Cushing, the chief executive officer, stated that QCoefficient can consistently deliver electricity bill savings of 30\% or more. 

Taken as a whole, the IBO session on today's business offerings showed that a number of newer or smaller companies now offer products with compelling features. As we discuss further in Section \ref{futureDirections}, example features include streamlining data access and acquisition, reducing control system deployment effort through automated workflows, optimizing for objectives beyond energy cost savings, and providing an ecosystem of building services beyond control, such as commissioning, fault detection and diagnostics, and guiding design and retrofit decisions. However, most of these companies are relatively early-stage; much remains to be seen about their profitability and the long-term sustainability of their business models. In particular, while several companies claim to have demonstrated clear utility bill savings, information about deployment costs and effort remains sparse. Early-stage start-ups may not have the resources to do frequent or large-scale field studies, or may not yet have moved their products down the cost curve. Disclosing deployment costs may also undercut start-ups’ abilities to charge competitive prices for their services.

\section{State-Of-The-Art Field Demonstrations}
\label{fieldDemonstrations}

In parallel with the companies that are innovating in the field of advanced commercial HVAC control, researchers in universities and federal laboratories continue to develop new algorithms and control paradigms. While this research historically happened mainly in simulation, a growing number of groups have shifted to real-world demonstrations of new technologies \cite{khabbazi}. To give a flavor for this field work, and to evaluate how it might shape future business offerings, the 2023 IBO workshop featured presentations from four cutting-edge field demonstration projects.

Hanmin Cai from the Swiss Federal Laboratory for Materials Science and Technology (EMPA) presented a field deployment of Data-Enabled Predictive Control (DeePC) \cite{yin2024data}. By contrast to conventional MPC, which typically involves modeling or system identification, DeePC directly uses input-output data from the system to optimize control actions using insights from behavioral systems theory. The field demonstrations of DeePC for space heating, water heating, and battery operation at the NEST facility on EMPA's campus showed robust constraint satisfaction and significant cost savings. The DeePC approach appears to eliminate the need for building modeling, a well-established hurdle to deploying MPC in real buildings \cite{sturzenegger2015model}.

David Blum from Lawrence Berkeley National Laboratory (LBNL) presented field demonstrations of the open-source MPCPy platform to reduce energy use by rooftop units (RTUs) that condition an office building on LBNL's campus \cite{blum2022field, kim2022site}. Blum's team modeled each zone as a second-order thermal circuit, used polynomial curve fits to represent fan and compressor powers, and adjusted RTU supply air temperature and fan speed set-points. Over 31 test days during the shoulder season in the mild Northern California climate, MPC reduced RTU energy by about 40\% while maintaining satisfactory thermal comfort. LBNL facilities engineers subsequently programmed reset schedules that emulate MPC behavior and achieve much of its energy savings with far less complexity.

Helia Zandi from Oak Ridge National Laboratory (ORNL) presented on a collaboration with Southern Company (a large utility), the University of Tennessee at Knoxville, and several companies \cite{tsybina2023findings}. The field demonstrations involved supervisory MPC and RLC of space and water thermostat set-points for homes in Atlanta, Georgia, and Birmingham, Alabama. MPC and RLC performed similarly, reducing heating and cooling energy costs by about 30\%, water heating energy costs by 18--39\%, and peak demand by about 15\%. 

Austin Lin and Johanna Mathieu from the University of Michigan discussed five years of field experiments on shifting commercial HVAC electricity use to provide wholesale ancillary services \cite{de2024living}. Example ancillary services include frequency regulation (quickly ramping demand up or down to regulate the alternating current waveform at its nominal frequency) and operating reserves (the capacity to quickly curtail demand in response to contingencies such as power line failures or generator outages). They shifted load over sub-hourly time scales by sending a single air temperature setpoint adjustment signal to most of the building's thermal zones. Lin and Mathieu showed that fast load shifting to provide wholesale ancillary services is technically feasible, but discovered an unexpected tension between load shifting and energy efficiency. They concluded that ancillary service provision likely makes economic sense in many circumstances, but some stakeholders may need additional compensation to cover potentially increased energy use and equipment wear and tear.

Asked to discuss implementation challenges, the presenters brought up some recurring themes. Establishing good working relationships with building operators and occupants takes time and persistent effort. Gaining and maintaining reliable, secure access to data and control privileges is another ongoing problem, and perhaps one best solved by subcontracting with companies that specialize in data services. Several presenters reported that addressing these practical challenges took far more effort than actually developing and implementing advanced control algorithms. We discuss these issues in more depth in Section \ref{futureDirections}.

Taken as a whole, the session on state-of-the-art field demonstrations showed that recent advances in algorithms and open-source software tools have significant potential to increase the revenue side of the business case for advanced commercial HVAC control systems. However, while all presenters qualitatively discussed deployment challenges, there was comparatively little effort devoted to quantifying deployment costs, which determine the other side of the business case. We believe that characterizing and mitigating deployment costs are significant gaps in today's research practice, as demonstrating a convincing case for advanced commercial HVAC control will require cost-benefit analysis. 

While the field demonstrations discussed above were bespoke undertakings, we expect that the advanced commercial HVAC control field can outgrow the current fragility through scale implementation, enabled through channel partners, productization, and efficient data access and system integration. Apart from these advances, the workshop presentations showed that relationship-building to convince stakeholders of the value of advanced controls is both necessary and potentially onerous. We discuss this further in Section \ref{businessConsderations}.

\section{Challenges and Opportunities}
\label{futureDirections}

After studying current business offerings and state-of-the-art field demonstrations, the 2023 IBO workshop concluded with an interview with James Dice, founder and chief executive officer of Nexus Labs, followed by a roundtable discussion among an audience of experts on the challenges and opportunities ahead. The interview and discussion touched on four major themes: Business considerations, data access and system integration, benefit/cost analysis, and truth in advertising.

\subsection{Business Considerations}
\label{businessConsderations}

\paragraph{Relationships:} As voiced by Vince Cushing from QCoefficient, David Blum from LBNL, and Babak Mohajer from Belimo, gaining the data access and control privileges required to deploy advanced HVAC control typically entails building relationships and trust with building owners, operators, and occupants. All of this takes time. To support trust-building, researchers and companies can do well-executed field studies with rock-solid measurement and verification (M\&V, discussed further in Section \ref{truthInAds}) procedures, widely publicized in language that non-experts can readily understand. These studies should help convince stakeholders of the reliability and benefits of advanced HVAC control.

\paragraph{Painless savings:} As voiced, for example, by James Dice from Nexus Labs, the community ought to consistently demonstrate ``painless savings.'' We cannot inconvenience occupants or annoy building operators. Even a few occupant call-backs or a frustrated building operator are enough to kill a project, no matter how promising the savings. These issues have contributed substantially to a widespread perception among decision-makers in businesses and the Department of Energy that savings from advanced HVAC control are fragile.

\paragraph{Broken promises:} Especially in large commercial buildings, the idea of ``optimizing'' building operations is not new. Building operators will almost certainly have heard promises to ``optimize'' (not necessarily in a rigorous mathematical fashion) from previous vendors, many of whom did not deliver on those promises. As voiced, for example, by Vince Cushing, these ``sins of past companies'' have ``poisoned the well,'' such that new companies, even those with solid business offerings based on reliable technologies, often encounter high barriers to adoption. As a remedy, companies could perhaps avoid using overloaded or inflationary terms such as ``optimization'' and ``artificial intelligence.''

\paragraph{Scale channel partners:} In the interest of scaling advanced HVAC control to many buildings, once a company has an economically viable product, it may make sense to find a scale channel partner, i.e. to collaborate with another company that has existing relationships with large numbers of building owners and operators. This could enable a new company to avoid the unsustainable costs associated with marketing, sales, and building trust with occupants and building operators one building at a time.

\paragraph{Productization:} Building owners are not interested in buying {\it algorithms}; they want {\it whole products} that include installation, commissioning, maintenance, support, and updates. For example, HVAC equipment manufactured with the on-board capability to provide ancillary services (such as those investigated by Austin Lin and Johanna Mathieu at the University of Michigan) might resonate more with building owners than an unfamiliar software solution provided by a third party. As James Dice put it, ``building operators do not want to think about any of the [data and control challenges] we're talking about here today.''

\subsection{Data Access and System Integration}
\label{dataAccess}

\paragraph{The data stack:} Deploying an advanced commercial HVAC control system involves three infrastructure layers: (1) device, (2) network, and (3) data. In the device layer, each device must have functional sensors and/or actuators and the ability to transmit and/or receive data. In the network layer, each device must have a reliable connection that can handle the required data rates. In the data layer, applications must reliably process, store, and access data. Rather than trying to establish or maintain all three data layers themselves, companies that seek to deploy advanced HVAC control may be well-served by letting other companies manage some or all of the system integration challenges. For example, if the appropriate data-sharing agreements were in place, the Belimo Energy Valve could provide valuable data that would otherwise be unavailable or require time-consuming integration with BAS. For another example, a start-up called Enode sells, as a subscription-based service, integration and maintenance of application programming interface (API) access to distributed energy resources. While a company that only wants to do control may view dependence on another company's middleware for data services as a liability, a building operator that wants numerous data-driven services, such as fault detection and diagnostics or tenant submetering, may prefer to have a single entity take responsibility for the data layer.

\paragraph{Path of least resistance:} Along the same lines, control companies should look for application areas where the device, network, and data challenges are easier or have already been solved. For example, small residential and commercial buildings usually have much simpler systems than large commercial buildings. Often, advanced residential HVAC control only requires integration with a smart thermostat. For another example, most mid-sized commercial buildings use RTUs, which tend to be standardized across large swaths of the market. This suggests an opportunity to create replicable, turnkey solutions for certain building classes, characterized by similar equipment and field conditions, rather than requiring unique field customization for each deployment. As alluring as large commercial buildings may be -- particularly the largest 3\% that actually have BAS today -- the unique system integration challenges and business barriers in this market segment may make other application areas more promising. We note, however, that while data challenges may be easier in small residential or commercial buildings, these market segments bring their own challenges, such as more variable occupant behavior, lower thermal mass, a range of privacy concerns, and sparser measurements. Helia Zandi from ORNL and Justin Hill from Southern Company discussed these issues in the context of their residential community field study.

\subsection{Mandate for New Business Thinking}
\label{businessThinking}

\paragraph{Outgrowing fragility:} The advanced commercial HVAC control community needs to overcome the perception among decision-makers in businesses and government, as articulated by workshop participants from investor-owned utilities and federal government agencies, that savings from advanced HVAC control are fragile. Unlike improved insulation or more efficient heat pumps, which continually save energy and reduce demand peaks, performance improvements from control can evaporate if any link in the data stack breaks down. For example, a network connection could fail or a frustrated building operator could pull the plug. To begin overcoming this perception of fragility, we need more case studies of substantial,  persistent savings in real buildings, demonstrated through trustworthy M\&V.
    
\paragraph{Beyond savings:} In addition to demonstrating reliable savings in the real world, the community must also answer a harder question: Can savings be achieved in an economically sustainable way that a business would actually want to pursue? In this vein, companies such as BuilTwins and Passive Logic have begun to bundle advanced HVAC control with other services that customers may value, such as fault detection and diagnostics and supporting design and retrofit decisions.

\paragraph{Deploying within a budget:} We should deploy technologies quickly, at scale in real testbeds, and within realistic budgets enabled by soundly demonstrated cost and carbon savings. We should track and report our deployment labor and costs. This raises a research opportunity to develop challenge problems with limited time and hard constraints on deployment effort. Workshop participants highlighted several innovative opportunities to compress deployment effort. For example, Passive Logic uses automated workflows to expedite the process of on-boarding a new building. For another example, Hanmin Cai from EMPA presented the DeePC methodology, which can reduce or eliminate the need for thermal modeling before deploying advanced HVAC control. (We note, however, that DeePC methods may need further work to demonstrate suitability for deployment at scale, particularly when applied to nonlinear HVAC systems.) David Blum from LBNL also mentioned the intriguing possibility of designing good rule-based controllers that mimic MPC or RLC behavior, achieving similar economic benefits with much less deployment effort. This idea ties in with past work on rule extraction from MPC, such as \cite{may2011model}.
    
\paragraph{Emphasizing additional benefits:} While energy cost savings are important, we should not lose sight of the additional benefits that advanced supervisory control can provide, such as improving occupant comfort, future-proofing against increasingly complex utility rate structures, detecting and diagnosing equipment faults, and reducing maintenance costs. For example, Austin Lin and Johanna Mathieu demonstrated that advanced HVAC control can provide new revenue streams by allowing building owners to sell ancillary service in wholesale electricity markets. (We note, however, that Lin and Mathieu also highlighted trade-offs between ancillary service revenues and increased energy use and equipment wear and tear. Similar trade-offs may also arise between other objectives of advanced control.) For another example, Vince Cushing emphasized that QCoefficient provides value to customers not only by improving energy efficiency, but also through coincident peak demand charge reductions and compliance with greenhouse gas regulations, such as New York City's Local Law 97. Cushing noted that, in his experience, many New York City building operators would rather reduce greenhouse gas emissions through advanced control than through hardware-based approaches, which are typically more capital-intensive and intrusive to occupants.

\subsection{Truth in Advertising}
\label{truthInAds}

\paragraph{Re-commissioning before control:} Savings depend as much on the quality of the benchmark control method as on the quality of the new, advanced control method. David Blum from LBNL, Filip Jorissen from BuilTwins, and several other workshop participants brought up this recurring theme. A large body of research has shown that significant savings are achievable in many existing buildings simply by re-commissioning systems and resolving long-standing faults, such as stuck dampers or clogged filters \cite{katipamula2005methods}. This process should come before implementing advanced control; otherwise, proponents of advanced control may take credit for savings that could be obtained in simpler ways. It behooves us as researchers to clearly delineate savings due to re-commissioning from savings due to advanced control. If we don't do this, then we as a research community may justifiably be accused of double-counting. While a company may provide value both through re-commissioning and advanced control, researchers should clearly differentiate between the two causes for savings. 


\paragraph{Toward community standards:} M\&V should ideally be done by neutral third parties, not researchers with vested interests in publishing high-profile papers with attractive, attention-grabbing results. The logistics of neutral M\&V could resemble those of single- or double-blind peer review. This raises a research opportunity to develop open-source methods and standards for measuring and verifying savings from field experiments, as well as research community buy-in to using these methods when possible, to the benefit of all involved. For example, a recent survey of advanced residential HVAC control studies revealed that savings reported relative to {\it modeled} counterfactuals, rather than counterfactuals based on {\it field-deployed} baseline control strategies, estimated consistently higher savings \cite{pergantis2024field}; this result suggests that there may be inherent biases in savings estimates depending on the types of baselines adopted. Discrepancies between modeled and measured performance could stem from model mismatch or from the inability to have control over all devices and systems. Suggested M\&V standards may require the more conservative baseline approach and explicitly advise against using simulation models to estimate savings.

\subsection{The Promise Remains}

\paragraph{Growing operational complexity brings new opportunities\dots} The difference between good heuristics and advanced control gets larger when incentives get more complex. We as a community should test advanced control under emerging incentives such as peak shaving, annual capacity contribution charges,  time-varying carbon intensities, emergency response to respiratory pathogens, more dynamic occupancy patterns in the post-pandemic era, complying with local decarbonization laws, maintaining healthy indoor air despite wildfire smoke, etc. In these complex multi-objective settings, humans will have great difficulty designing good heuristics. As these settings become increasingly prominent, opportunities for advanced control increase as well.

\paragraph{\dots and also new risks:} Many regions are seeing increased adoption of electric vehicles, heat pumps, rooftop solar photovoltaics, and energy storage, all of which plug into buildings. As equipment configurations become increasingly complex, building operators may find that they {\it need} an advanced control system simply to ensure reliable and cost-effective operations. While rule-based controls could theoretically orchestrate these new devices, such orchestration is expensive and requires specialized domain expertise. Advanced supervisory control algorithms such as MPC, DeePC, and RLC may be the only viable options to accommodate such complexity at scale.

\subsection{Main Findings of the Session}

Taken as a whole, the roundtable discussion showed that many of the barriers to adoption of advanced commercial HVAC control are not fundamentally technological. For example, significant challenges arise from managing relationships with incumbent firms and with the people who operate and occupy buildings. Past companies' failures to deliver on their promises also complicate marketing and sales. 

The roundtable discussion also showed that of the barriers that {\it are} technological, many are not algorithmic in nature. For example, companies often encounter significant challenges in gaining access to historical data, real-time sensor readings, and control authority over actuator set-points. Nevertheless, algorithmic barriers to adoption do exist and present opportunities for researchers to meaningfully accelerate technology adoption. For example, researchers could design algorithms that need far fewer sensors and actuators, so require much less integration effort. For another example, researchers could design algorithms for emerging, complex incentives such as reducing power demand during system-wide peaks, responding to real-time carbon intensities, quickly adapting to reports of respiratory pathogens or to spikes in wildfire smoke intensity, and complying with greenhouse gas emission regulations such as New York City's Local Law 97. For a third example, researchers could design algorithms that coordinate the operation of HVAC equipment alongside electric vehicles, stationary energy storage, or solar photovoltaics.

\section{Conclusion}
\label{conclusion}

This paper summarized findings from the 2023 Intelligent Building Operations Workshop. It presented case studies of three emerging business offerings; reviewed six technology demonstrations from universities and government research laboratories; and highlighted key themes from a roundtable discussion among experts. These sessions showed that the potential of advanced commercial HVAC control has not yet been realized at scale; that current research and development efforts show great promise but also underemphasize important business considerations such as deployment effort and costs; and that research opportunities exist to meaningfully accelerate adoption by developing algorithms that deploy at very low cost, optimize a range of existing and emerging economic incentives, and harmonize HVAC operation alongside new energy equipment such as electric vehicles, stationary energy storage, and solar photovoltaics. 

In light of the workshop's findings, we argue that our research community should refocus its efforts on characterizing and improving the business case for advanced commercial HVAC control systems. Specifically, we recommend that researchers, funding agencies, journal editors, and reviewers of papers and proposals should emphasize the following:
\begin{enumerate}
    \item Internalize lessons learned from past field studies of advanced commercial HVAC control.
    
    \item Prioritize field studies over simulations, especially field studies that explore new control objectives or that coordinate HVAC equipment with electric vehicles, energy storage, or solar photovoltaics.
    
    \item Optimize and fully disclose not only economic benefits, but also overall deployment effort and costs, including ongoing maintenance.

    \item Investigate how benefits and costs depend on building characteristics, climate, control system architecture, and control objectives. This should identify economic ``sweet spots,'' where benefits are large relative to deployment costs, and where the advanced commercial HVAC control industry could scale up buoyed by successes.  
    
\end{enumerate}

Finally, although the workshop did not focus on policy, we encourage researchers to engage with the policymakers who shape the economic and legal environments in which advanced commercial HVAC control systems function. Historically, good technologies have not always prevailed over inferior incumbents; good technology needs good policy as an ally. In particular, international collaboration on policymaking can foster standardization and consistent rule-making. For example, international collaboration has recently led to standards for energy flexibility in buildings resulting from HVAC control through International Energy Agency Annexes 67 (Energy Flexible Buildings \cite{iea67}), 81 (Data Driven Smart Buildings \cite{iea81}), and 82 (Energy Flexible Buildings Towards Resilient Low Carbon Energy Systems \cite{iea82}). The International Energy Agency is currently guiding a proposal for a future project, ``Grid Integrated Control of Buildings.'' This international collaboration will emphasize the development of a digitalization framework for integrating flexible resources with grid markets, practical demonstrations to increase technical readiness of flexible demand, and investigating the impacts and opportunities for managing real-time emissions. Researchers who engage with these efforts could help shape the fundamental conditions that influence the demand for, and success of, advanced commercial HVAC control systems.

\appendix

\section{Summaries of Current Business Offerings}

\subsection{BuilTwins' White-Box Model Predictive Control}

Filip Jorissen is the co-founder and Chief Technical Officer of BuilTwins \cite{builtwins}, a start-up that spun out of Lieve Helsen's research group at KU Leuven in Belgium. Jorissen presented on BuilTwins' MPC system that uses detailed, physics-based models of buildings and HVAC components \cite{jorissen2022white}. BuilTwins offers a range of services including energy system design assistance, commissioning, monitoring, periodic reporting, and control. For control, BuilTwins models every component in the thermal and mechanical systems in Modelica using nonlinear differential-algebraic equations and the IDEAS library \cite{wetter2014modelica}. The controller model is constructed through a web-based graphical user interface, which is automatically mapped into a Modelica model, which is automatically translated into a nonlinear optimization problem to be solved in real time \cite{jorissen2021strengths}. To handle situations where some necessary data are missing, such as older buildings with outdated mechanical schematics, BuilTwins is developing an automated calibration method that uses a limited amount of sensor data. This semi-automated toolchain for model and MPC formulation, starting from HVAC hydraulic schemes and architectural plans, significantly reduces the expertise needed. This modeling approach has notorious speed limitations. BuilTwins has overcome these limitations by developing custom solvers, through years of research at KU Leuven, that are fast enough for real-time control.

On the business side, BuilTwins only recently launched. While initial uptake has been promising, they cannot yet comment on challenges associated with scaling up the business. Jorissen brought up what would become a recurring theme in the workshop: ``The main driver of savings is how bad the original control was, not how good you’re doing.''

\subsection{Belimo Automation's Energy Valve}

Babak Mohajer, research engineer at Belimo Automation \cite{belimo}, presented on the evolution of Belimo's Energy Valve. The Energy Valve is a flow control device for hydronic heating and cooling applications that consists of a valve, a flow meter, and supply and return water temperature sensors. The Energy Valve can detect when the temperature difference between the supply and return water streams becomes too small, indicating that the heat exchange device is producing marginal thermal output for the water-side flow rate. By constraining flow at the heat exchanger level, the Energy Valve avoids operation of the heat exchanger in the ``energy waste'' zone indicative of decreasing marginal thermal power per unit increase in flow as well as contributes to the maintenance of a large temperature differential at the system level. 

In addition to the Energy Valve hardware, Belimo provides a suite of analytical tools that can turn measurements from the Energy Valve into actionable information. Mohajer presented the three evolutionary stages of the Energy Valve's data analytics: (1) an Excel tool, (2) a human-in-the-loop support system, and (3) a fully automated, machine-learning-based system.

On the business side, Mohajer reported that the buildings industry is slow to adopt new technologies. Belimo had to expend significant effort to bring their product to market at scale. In particular, Belimo had to develop a ``product push'' rather than ``customer pull'' strategy; they had to convince each new customer that the Energy Valve could provide value. This required investment in pilot projects, field demonstrations, and case studies. Mohajer also reported that the benefits to customers are highly variable and application-specific. Furthermore, Belimo bundles the advanced benefits, such as their data analytics, with the valve at no additional cost to the customer. Belimo has not had success offering these unique benefits as an add-on subscription.

\subsection{Passive Logic's Building Automation System}

Passive Logic \cite{passivelogic} has developed a BAS that is completely new, from both a hardware and a software perspective. Troy Harvey, Passive Logic's co-founder and Chief Executive Officer, argued that buildings are among ``the most complex cyber-physical systems'' in terms of spatial complexity and the numbers of decision variables, sensors, and actuators, and so are worthy of a completely new approach to automation. Passive Logic's approach involves a scalable hardware platform (the ``Hive''), of which a smaller building may have a few and large buildings may have many. Each Hive instance communicates with the others, forming a network of shared computing resources. On the hardware, Passive Logic runs a real-time version of ``Quantum,'' their digital twin software, which uses a graph representation of networked HVAC component models to optimize control setpoints. Operational data is used to identify component model parameters by automatic differentiation of a loss function through the graph, akin to back-propagation in neural network training. Harvey presented an ecosystem of software tools that enable control system design, deployment, commissioning, and monitoring \cite{harvey2022autonomous}. 

Unlike Belimo, Passive Logic reported that their technology offering enjoys a ``customer pull'' rather than relying on a ``technology push.'' However, Harvey reported that, based on his experience and conversations with others in the industry, it is currently difficult to attract talent to the BAS industry, where dated software tools abound and, in Harvey's view, the last big innovation (Tridium's Niagara JACE platform) happened a quarter-century ago. Harvey argued that Passive Logic's new ecosystem of software tools are game-changers in making building control an attractive area of employment for young talent interested in technology.

\section{Summaries of Field Demonstrations}

\subsection{Data-Enabled Predictive Control at EMPA}

Hanmin Cai, scientist at EMPA in D{\"u}bendorf, Switzerland, presented on EMPA's field deployment of Data-Enabled Predictive Control (DeePC). By contrast to conventional MPC, which typically involves a modeling or system identification phase, DeePC directly uses input-output data from the system to optimize control actions. DeePC uses results from behavioral systems theory, which allows nonparametric representation of an arbitrary linear system given a persistently exciting trajectory of inputs and outputs. Cai and collaborators developed an advancement to allow for measurement noise and probabilistic constraints on future system outputs, an approach referred to as Signal Matrix MPC \cite{yin2024data}.

Cai reported results from both simulations and field deployments at the NEST facility on the EMPA campus. The field results involved multiple hardware applications, including space heating, water heating, and battery operation. Cai reported that their modified Signal Matrix MPC algorithm not only performed well in the test cell in which they tuned algorithmic hyperparameters, but also transferred well to new test cells with no additional hyperparameter tuning. The algorithm has fewer hyperparameters than many data-driven approaches, which facilitates practical implementation.

In terms of practical deployment considerations, Cai remarked that their algorithm respects thermal comfort constraints much more robustly than any competing algorithm. For this reason, the Signal Matrix MPC approach should promote occupant satisfaction throughout the deployment and commissioning phases. Cai also mentioned that this is a black-box technique, so it suffers from similar explainability deficits to RLC. This could complicate applications that integrate a large number of grid-interactive buildings into critical infrastructure, where explainability is key to ensuring safe system operation and timely diagnosis and recovery from interrupted operation. Finally, Cai reported that although the underlying behavioral systems theory results apply only to deterministic linear systems, the methods appear to perform well even for stochastic or nonlinear systems.

\subsection{Model Predictive Control at LBNL}

David Blum, a research scientist at Lawrence Berkeley National Laboratory (LBNL), presented on field demonstrations of MPC in two stories of an office building on LBNL's campus, covering 60,000 ft$^2$ of conditioned floor area \cite{blum2022field, kim2022site}. A separate data center in the same building was not part of the control study. Four typical rooftop units (RTUs) supply conditioned air to under-floor air distributions systems with reheat. Zone temperature control is maintained in the perimeter zones by under-floor variable air volume terminal boxes. The predictive controller seeks to minimize RTU energy use while maintaining comfortable zone air temperatures. From the building under control, a range of systems including HVAC, electrical, and on-site weather measurements, communicate with a remote server. This remote server collects historical data, forecasts weather and internal loads, identifies model parameters, and performs state estimation and control optimization. The server runs the open-source MPCPy platform, represents each zone's thermal dynamics using a second-order linear thermal circuit model, and uses polynomial curve fits for fan and compressor powers. Internal loads are estimated from circuit-level submeter measurements. The server sends supervisory control set-points of supply air temperature and fan speed to each RTU, where a watchdog system screens the set-points before passing them to the RTU's on-board control system. 

Over 31 days of MPC trials, the LBNL team estimated that MPC reduced RTU input energy by about 40\% while maintaining similar thermal comfort to the prior control system. These trials occurred between October and December in Berkeley, California, when weather conditions were generally mild. Savings were largest under moderate outdoor air temperatures, and decreased in both hot and cold weather. In future work, the team plans to quantify MPC savings during both winter and summer. Blum reported that LBNL facilities engineers subsequently programmed reset schedules that emulate MPC behavior by adjusting supply air temperatures based on the outdoor temperature. Blum also reported promising preliminary results involving MPC for electricity price arbitrage, rather than energy efficiency. 

In terms of practical deployment considerations, Blum -- like Jorissen -- reported that savings depend more on how bad the pre-existing control system was rather than on how well MPC performs. Blum emphasized that coordinating with building operators takes time and effort, but is necessary to establish trust before gaining access to secure systems, to build understanding of existing system operation, to edit logic in building management systems, and to respond to any faults that arise. The LBNL team used an array of alarms and watchdogs to handle network failures, service interruptions, sensor fouling, and other faults.

\subsection{Model Predictive Control at QCoefficient}

Vince Cushing, co-founder and Chief Technical Operator of QCoefficient, summarized his experience with a number of field implementations. At a high level, QCoefficient optimizes zone air temperature setpoints using high-fidelity EnergyPlus models, with a focus on interactions between HVAC equipment and the power grid. Rather than continuous closed-loop control, as one implementation, QCoefficient optimizes set-points once per day to minimize information traffic. A building operator screens the set-points for acceptability before sending them to the BAS via an interoperability gateway (Tridium's Niagara JACE controller). The once-per-day optimization is not an inherent limitation of QCoefficient's systems, but a choice that makes cybersecurity management much easier and builds trust with building operators as they can review the planned strategies. This approach has proven particularly effective in QCoefficient's work with federal buildings, whose automation systems reside behind a government firewall. QCoefficient has also worked with several large investment banks in New York City. QCoefficient's systems can optimize for a number of objectives, including energy efficiency, energy costs, peak demand, greenhouse gas emissions, and indoor air quality, either individually, as a blended cost function, or through multi-objective optimization. Through the field implementations, QCoefficient learned that even a small amount of building thermal mass activation can unlock primary and secondary HVAC system efficiency gains, for example by avoiding dispatch of older, less efficient chillers or harnessing affinity laws. Another area ripe for improvement is avoiding over-ventilation during time periods with low occupancy and high sensible or latent ventilation loads.

Cushing presented results from several field demonstrations in large commercial buildings, including the Willis Tower in Chicago (4 million ft$^2$ with low-efficiency equipment and envelope), a 2.3 million ft$^2$ corporate headquarters, and a 1.2 million ft$^2$ bank headquarters in New York City. A recurring theme across QCoefficient's case studies was that a deep understanding of commercial utility bills -- particularly incentives such as hourly electricity prices, peak demand charges, and annual capacity charges -- informs where the money-saving opportunities lie. In the Chicago case study, for example, Willis Tower often saw negative wholesale electricity prices overnight due to an abundance of wind power from Indiana and inflexible coal base-load generation. Building operators get paid to precool during these periods of negative prices. In the New York City case study, reducing demand during system-level peaks greatly reduced annual capacity charges. Cushing claimed that QCoefficient can reliably take 30\% off of annual electric bills. In New York City, however, utility bill savings are currently less of a motivator than greenhouse gas emissions due to Local Law 97, which places binding limits on emissions from buildings over 25,000 ft$^2$.

In terms of practical considerations, Cushing emphasized the importance of collaborating with large-scale partners that already have relationships with building operators through other product offerings. For example, Cushing mentioned that Johnson Controls, Siemens, and Schneider Electric each have about a one-quarter share of the BAS market, and they are protective of ``their'' buildings. Tapping into these large market shares provides an exciting opportunity for a new company to quickly scale up their operations. To collaborate well with large BAS providers, Cushing recommended providing complementary services, rather than competing with existing offerings. More pragmatically, changing set-points in a large commercial building requires access to the BAS, which must be obtained from the BAS provider. While this might only cost the provider around \$10,000 in internal labor costs, Cushing reported that QCoefficient typically pays around \$60,000 for the service.

Cushing reported that commercial building operators who seek to improve efficiency often prefer control upgrades, which come at lower capital costs, to the more expensive option of upgrading equipment or the building envelope. He also mentioned that it is not necessary to control the entire conditioned floor area of a building; often, controlling a few ``problem floors'' can build trust with building operators by achieving significant savings while improving occupant satisfaction.

Cushing argued that a significant impediment to gaining access to a commercial building, in terms of encountering stakeholders that are open to and interested in innovative building control approaches, is the fact that the stakeholders have likely had experiences with previous offerings, many of which did not deliver on their promises. New companies are thus held accountable for, or are at least scrutinized for, the failures of past companies. Critically, a new company must be able to convincingly answer question such as ``What makes you different?'' and ``How will you avoid the problems of the previous company?''

\subsection{Smart Neighborhoods from Southern Company}

Helia Zandi from Oak Ridge National Laboratory (ORNL) presented on two field demonstrations, led by Southern Company in collaboration with several industry partners and the University of Tennessee at Knoxville \cite{tsybina2023findings}. The field demonstrations involved residential communities of 46 townhouses in Atlanta, Georgia, and 62 single-family homes in Birmingham, Alabama, as well as two individual test houses operated by ORNL and Southern Company. Zandi and her colleagues deployed both MPC and RLC for space conditioning and water heating at ORNL's Yarnell Station Research House, and MPC in the Atlanta and Birmingham communities. Both control approaches used similar sensing and communication infrastructure, consisting of cloud platforms for prediction, aggregate-level optimization, and building-level optimization. The teams installed a range of sensors, including occupancy detectors, and communicated with smart thermostats via APIs over WiFi. Building-level controllers decided indoor air temperature set-points and hot water set-points, which the equipment manufacturers' device-level controllers tracked. 

Zandi reported that both MPC and RLC substantially reduced energy costs and demand peaks. In both field demonstrations and simulations, supervisory control reduced energy costs for heating and cooling by about 30\%, and for water heating by 18--39\%. For water heating, RLC reduced energy costs by 1--3\% more than MPC, surprisingly. Supervisory control also reduced peak demand for space conditioning by 75--140 W per home (about 15\%), and peak demand for water heating by 40--65 W per home (about 60\%).

In terms of practical implementation challenges, Zandi mentioned difficulties with installing and commissioning sensors and with integrating diverse equipment into a unified data platform. Even after the initial integration phase, equipment manufacturers occasionally altered data formats via firmware updates or API changes with short notice. Accommodating these data format changes required ongoing engineering effort. Zandi also mentioned that user education, buy-in, and satisfaction are critical. In particular, users may worry about privacy or cyber-security issues related to their home usage patterns. Furthermore, the structure of electricity pricing strongly influences the potential savings from advanced control. Justin Hill, Zandi's collaborator at Southern Company, mentioned that off-the-shelf technologies can be rather fragile, with ``unfamiliar installation challenges'' and a ``lack of operational understanding.'' 

From the utility's perspective, Hill reported a general lack of understanding of the real-world value of grid-interactive efficient buildings (GEBs), or even how to value GEBs. For a utility, activating the demand response potential of an aggregation of GEBs, possibly with humans in the loop, is much more complex and less trusted than simply building new distribution or generation infrastructure. Hill also raised legal and privacy concerns and cited persistent difficulties with fixing communication and data integration issues pertaining to firewalls, access permissions, API integration, and the need to await human actions when problems occur. Specifically, residential buildings have a wide range of proprietary or vendor-specific systems. The need to integrate these diverse systems limits scalability and the potential for portfolio-level operations.

\subsection{Fast Load Shifting at the University of Michigan}

Austin Lin and Johanna Mathieu from the University of Michigan presented on five years of field experiments on shifting commercial HVAC electricity use over sub-hourly time scales \cite{de2024living}. Their team worked with 14 campus buildings at the University of Michigan and four at North Carolina State University. The team shifted HVAC electricity use by adjusting zone air temperature setpoints (cooling setpoints only in Michigan; both heating and cooling setpoints in North Carolina). The team sent the same setpoint adjustments to all of the controllable zones in each building. The Michigan buildings used a Siemens BAS, while the North Carolina buildings used a Metasys BAS by Johnson Controls. 

The team primarily sought to understand the trade-offs between energy efficiency and fast load shifting. To investigate these trade-offs, they sent one-hour-periodic square wave setpoint adjustment signals, then observed the resulting perturbations in measured electricity use. These open-loop experiments resemble control actions that might arise when providing ancillary services, such as frequency regulation or spinning reserve, to a power system operator. The team found that these square-wave setpoint adjustments increased cumulative electricity use under some circumstances and decreased electricity use under other circumstances. Ventilation systems with faster response times appeared to have round-trip efficiencies, measured over full cycles of square-wave setpoint adjustments, that were closer to unity. The team concluded that practitioners should pay careful attention to the energy efficiency impacts of fast load shifting for ancillary services. Ancillary service provision likely makes economic sense in many circumstances, but some stakeholders may need additional compensation to cover potentially increased energy use. According to Lin and Mathieu, ``incentives need to align with maximizing the overall value of building-grid interaction.''

In terms of practical implementation challenges, Lin and Mathieu noted that the ease or difficulty of interfacing with a BAS depended more on the building operators than on the BAS brand or type. In North Carolina, for example, a building operator helped the team communicate with all buildings' BAS via a single Python instance through the open-source Bacpypes library. While this approach worked well, it required coordination with campus information technologies staff to navigate. Lin and Mathieu also noted that some building managers or occupants were protective of their spaces and preferred not to change HVAC equipment operation; the team excluded sensitive spaces from their experiments for this reason. Maintenance activities, such as cleaning filters in ventilation systems, sometimes affected experiments in unexpected ways. The team also found the electrical measurements in today's BAS insufficient for measuring fast load perturbations from setpoint adjustments. They feel that installing submeters to measure power on individual circuits will be necessary to meet the high reliability requirements that power system operators require from ancillary service providers.

\subsection{Open-Source Software Tools From PNNL}

Jan Drgona from Pacific Northwest National Laboratory (PNNL) presented on NeuroMANCER, a collection of open-source software tools for data-driven control of physical systems, demonstrated for the case of commercial HVAC systems \cite{drgovna2022differentiable, drgovna2022learning}. NeuroMANCER combines a system identification step, wherein a physics-informed neural network is trained on trajectories from the real system to represent the nonlinear system dynamics, with a policy learning step, wherein a second neural network is trained to implement a feedback control policy that minimizes an MPC-type loss function, such as minimizing operating cost while observing comfort constraints. The algorithms in NeuroMANCER are quite general and can be applied to nonlinear dynamical systems beyond buildings.

In terms of practical considerations, NeuroMANCER eliminates many of the implementation challenges associated with MPC, such as model development, disturbance prediction, and solving optimization problems. Given a set of training data with sufficient information content, NeuroMANCER produces a feedback control policy that can be implemented automatically in real time.

\section*{Acknowledgments}

The authors thank the contributors to the 2023 Intelligent Building Operations Workshop for sharing their valuable experiences and insights, and for reviewing drafts of this consensus document. In particular, we thank Nick Clements (University of Colorado) for logistics support; the speakers: David Blum (LBNL), Hanmin Cai (EMPA), Vince Cushing (QCoefficient), James Dice (Nexus Labs), Jan Drgona (PNNL), Troy Harvey (Passive Logic), Filip Jorissen (Builtwins), Austin Lin and Johanna Mathieu (University of Michigan), Babak Mohajer (Belimo), Helia Zandi (ORNL); and the participants: Kyri Baker (University of Colorado), Sourav Dey (University of Colorado), Bryan Eisenhower (Carrier), Davide Fop (Politecnico di Torino), Samuel Fux (Belimo), Lieve Helsen (KU Leuven), Tianzhen Hong (LBNL), Nicholas Long (University of Colorado and NREL), Peter McKinney (Carrier), Stefan Mischler (Belimo), Ozan Baris Mulayim (Carnegie Mellon University), Zoltan Nagy (University of Texas at Austin), Greg Pavlak (Pennsylvania State University), Anand Krishnan Prakash (LBNL), Amir Roth (DOE Building Technologies Office), Soumik Sarkar (Iowa State University), and Ettore Zanetti (LBNL).

\section*{Data availability statement}

Data sharing is not applicable to this article as no new data were created or analyzed in this study.

\printbibliography

\end{document}